\documentstyle[12pt]{article}
\advance\textheight by \topskip
\def \sl  { / {\hskip - 0.27 cm  {\bf P}}}

\begin{document}

\begin{flushright}
SU-ITP-97-23\\

hep-th/9705056\\
May 8, 1997
\end{flushright}
\vspace{.5cm}

\begin{center}
\baselineskip=16pt

{\Large\bf  Covariant Quantization of D-branes} \\

\

\vskip 1 cm

{\bf  Renata Kallosh}

\vskip 1cm

{\em Physics Department, Stanford
University, Stanford, CA 94305-4060, USA\\
kallosh@physics.stanford.edu
}

\end{center}

\vskip 1 cm
\centerline{\bf ABSTRACT}
\vspace{-0.3cm}
\begin{quote}
We have found that $\kappa$-symmetry allows a covariant quantization provided
the
ground state of the theory is strictly massive.
For D-p-branes a Hamiltonian analysis is performed to explain the existence of
a
manifestly supersymmetric and Lorentz covariant description of the BPS states
of
the theory. The covariant quantization of the D-0-brane is presented as an
example.

\end{quote}
\normalsize

\baselineskip=16pt

\newpage
\section{Introduction}
Extended objects with global supersymmetry have local $\kappa$-symmetry. This
symmetry is difficult to quantize in Lorentz covariant gauges keeping finite
number of fields in the theory. A revival of
interest to $\kappa$-symmetric objects is due to the recent discovery of
D-p-branes \cite{Pol} and D-p-brane actions \cite{Ce1,Ag1,Be1,Hull}.

Before the choice of the gauge is made the problem with covariant quantization
of $\kappa$-symmetry can be seen as the impossibility to disentangle
covariantly
the combination of the first and second class fermionic constraints. In the
past the quantization of the superparticle, of the supersymmetric string, and
of the
supermembrane was performed only in the light-cone gauge for spinors
\begin{equation}
\Gamma^+ \theta =0 \ .
\end{equation}
In this gauge  $\kappa$-symmetry is fixed leaving only the second class
fermionic
constraints whose Poisson brackets  are invertible
 even for the massless ground state.  In covariant gauges the major problem is
that the constraints are not invertible.

Recently a covariant gauge  fixing $\kappa$-symmetry  of D-branes has been
discovered \cite{Ag1}. The fermionic gauge is of the form
\begin{equation}
\theta_2 =0 \ ,
\end{equation}
where $\theta_2
$ is one of the  chiral spinors of the 10d theory. Moreover, since there is a
duality between the D-1 brane and the fundamental IIB string,  a gauge-fixing
of the fundamental string in covariant gauge has been achieved  in \cite{Ag1}
by passing. Does it mean that the previous attempt to covariantly quantize the
Green-Schwarz string missed the point,  or something else happened? Is the
existence of covariant gauges a special property of D-p-branes only, or they
exist  for all p-branes? To address these issues one has to take into account
that the covariant quantization performed in \cite{Ag1} also used the so-called
static gauges $X^m = \sigma^m$ for fixing the bosonic reparametrization
symmetry.
The total picture of covariant quantization of $\kappa$-symmetry with the use
of static
gauges is difficult to analyze.

Here we will first switch to a Hamiltonian
form of the theory which will allow us to study the issues in quantization of
$\kappa$-symmetry before a choice of the reparametrization fixing gauge is
made.
We will analyze the quantization of $\kappa$-symmetric theories and we will
establish connection of the quantized theory with  $d=10$, $N=2$ supersymmetry
algebra
with central   extensions:
\begin{eqnarray}
\{ Q_{\alpha}  ,  \; Q_{\beta}  \} &=& 2 (C \Gamma^{m})_{\alpha \beta}
\Bigl({\bf
P}_{m} + {Q^{N}_m \over 2\pi \alpha'}\Bigr) \ , \\
\nonumber\\
\{ \tilde Q_\alpha  ,  \; \tilde Q_\beta  \} &=& 2 (C \Gamma^{m})_{\alpha
\beta} \Bigl({\bf P}_{m} - {Q^{N}_m \over 2\pi \alpha'}\Bigr)\ ,\\
\nonumber\\
\{ Q_\alpha ,  \; \tilde Q_\beta \} &=2 &  \sum_A  (C \Gamma^{A})_{\alpha
\beta}\, T_{(p)} {Q_{A}^R \over p!} \ .
\end{eqnarray}
We present here the supersymmetry algebra in the form given in
\cite{Pol}.
Here $Q^N$ and $Q^R$ are NS-NS and R-R charges and $A$ runs over
antisymmetrized products of gamma matrices. We will find the covariant mass
formula for the quantized D-p-branes.

The quantization of D-p-branes in static gauges \cite{Ag1}  leads to
complicated non-linear actions. In this paper we will perform a covariant
quantization of the D-0-brane which will
give a simple quadratic action.

\section{Main results}
We have found  that the covariant quantization of D-branes  is consistent and
that the ground state $|\Psi\rangle$ of the system has a non-vanishing mass
\begin{equation}
 M^2  |\Psi\rangle   =  - (\Gamma^m {\bf P}_m)^2  |\Psi\rangle =
 T^2 \left ( \det (G+{ \cal F})_{ab}+ P^a G_{ab} P^b\right )  |\Psi\rangle \ ,
\label{grmass}\end{equation}
\begin{equation}
 M^2  >0 \ .
\end{equation}
Here ${\bf P}_m$ is the momentum conjugate to the coordinate $X^m$ and the
eigenvalue of  the operator $- (\Gamma^m {\bf P}_m)^2 $  is given by the
positive definite expression $(T^2 \det (G+{ \cal F})_{ab}+ P^a G_{ab} P^b) $.
 $T$ is the D-brane tension, the index $a$ runs over the space components
of the brane, $G_{ab} , { \cal F}_{ab}
 $ are the space part of metric induced on the brane and the 2-form,
respectively,
and $P^a$ is the momentum conjugate to the vector field on the brane.  Both the
metric $G_{ab}
 $ and the 2-form $ { \cal F}_{ab}
 $ are manifestly supersymmetric under the N=1 part of the full N=2 global
supersymmetry.

 For example, for the covariantly quantized D-2-brane we find from
(\ref{grmass}) that the mass of
the ground state  is
\begin{equation}
M_{(D2)}^2 =  T^2 \{ \Pi^m , \Pi^n \}^2 + ({ \cal F}_{12})^2 + P^r G_{rs} P^s,
  \qquad   m,n = 0,1 \dots ,9; \quad r,s = 1,2.
\label{massD2}\end{equation}
Here the square of the Poisson bracket is defined as,
\begin{equation}
\{ \Pi^m , \Pi^n \}^2 \equiv [\left( X^m{}_{,r } - \bar \lambda \Gamma^m
\lambda_{ ,r } \right)
 \left( X^n{}_{,s } - \bar \lambda \Gamma^n \lambda_{ ,s } \right) -
(m\rightarrow n)]^2 \ , \end{equation}
and $\lambda$ is one of the  chiral d=10 spinors which remains in the theory
after $\kappa$-symmetry is gauge-fixed covariantly.
This is a Lorentz covariant  generalization of the  mass operator
\begin{equation}
M_{(M2)}^2 = \{X^i, X^j\}^2 - \bar \theta \Gamma_- \Gamma_i \{ \theta , X^i \},
 \qquad i,j
=1,\dots 9,
\end{equation}
which appears in the quantization of the M-2 supermembrane \cite{BST} in the
light-cone
gauge \cite{BSTa,DWHN}.
In the form in which $r, s$ are matrix indices this expression
has been used in M(atrix )model \cite{BFSS}.

The static gauge for the D-2 brane $X^1 = \sigma, X^2 = \rho$ corresponds to
\begin{equation}
\{ \Pi^1 , \Pi^2 \} =  (X^1)_{,\sigma}  \,  (X^2)_{, \rho}  + \dots = 1+\dots
\end{equation}
This makes the constant part of the square of the momentum on the ground
state
non-vanishing as long as the tension $T$ is non-vanishing.
Thus the covariant quantization of the D-2-brane performed in \cite{Ag1}
indeed confirms our general conclusion  that the ground state of a D-brane has
a non-vanishing mass as long as the tension of the D-brane is non-vanishing.

This observation also solves the apparent paradox with the covariant
quantization of $\kappa$-symmetry of the IIB fundamental string in \cite{Ag1}.
 For  the D-1-brane we get
\begin{equation}
M_{(D1)}^2 =  T^2 (\Pi^1)^2 + P^1 G_{11} P^1\ .
 \end{equation}
The ground state of the D-1-brane is massive. The technical reason for this (in
Lorentz covariant gauge for spinors and  the static gauge $X^1 = \sigma$ for
reparametrization symmetry \cite{Ag1})  is that $(\Pi^1)^2=[(X^1)_{, \sigma
}]^2 + \dots = 1 +\cdots$.
 This corresponds to the D-string wrapped around the circle.
In case of a fundamental string, the covariant gauge-fixing proposed in
\cite{Ag1}
also uses a static gauge. Therefore {\it the  massless state
of the fundamental GS string is projected out},  and in this way the two
quantized string theories, D-1-brane   and type IIB fundamental strings are
dual to each other.
For the D-0-brane the mass formula (\ref{grmass}) cannot be applied since there
are no space directions.  However, the mass formula in this case is  simply
$M^2=Z^2$, where $Z$ equals the tension of the D-0-brane.

 We will proceed with the derivation of the results stated above.

\section{Irreducible $\kappa$-symmetry on D-branes}

The $\kappa$-symmetric D-brane action in the flat background
geometry\footnote{We use
notation of \cite{Ag1}.} consists of the
Born-Infeld-Nambu-Goto term $S_1$ and Wess-Zumino term $S_2$:
\begin{equation}
S_{\rm DBI} +S_{\rm WZ} =T\left (  - \int d^{p + 1} \sigma\, \sqrt{- {\rm {\rm
det}} (G_{\mu\nu} + {\cal F}_{\mu\nu})} +\int \Omega_{p + 1} \right) \ .
\label{action}\end{equation}
Here $T$ is the tension of the D-brane, $G_{\mu\nu}$ is the  manifestly
supersymmetric induced world-volume metric
\begin{equation}
G_{\mu\nu} = \eta_{mn} \Pi_\mu^m \Pi_\nu^n \ , \qquad
\Pi_\mu^m = \partial_\mu X^m - \bar\theta \Gamma^m \partial_\mu \theta \ ,
\end{equation}
and ${\cal F}_{\mu\nu}$ is a manifestly supersymmetric Born-Infeld field
strength (for $p$ even)
\footnote{We define
 spinors  for even $p$ as $\theta = \theta_1 + \theta_2$ where  $\theta_1
\equiv
{1\over
2}(1+\Gamma_{11})\theta$
 and $\theta_2 \equiv {1\over 2}(1-\Gamma_{11})\theta$.}

\begin{equation}
{\cal F}_{\mu\nu} \equiv F_{\mu\nu} - b_{\mu\nu} =  \Bigl[\partial_\mu A_\nu -
\bar\theta \Gamma_{11} \Gamma_m \partial_{\mu}\theta
\Bigl(\partial_{\nu} X^m - {1\over 2} \bar
\theta \Gamma^m \partial_{\nu}\theta\Bigr)\Bigr] - (\mu \leftrightarrow \nu) \
{}.
\end{equation}
 When $p$ is odd, $\Gamma_{11}$ is replaced by
$\tau_3\otimes I$.  The action has
global supersymmetry
\begin{equation}
\delta_\epsilon \theta = \epsilon, \qquad \delta_\epsilon X^m = \bar\epsilon
\Gamma^m
\theta \ . \label{susytrans}
\end{equation}
and local $\kappa$-supersymmetry:
\begin{equation}
\delta X^m = \bar\theta \Gamma^m \delta\theta = - \delta \bar\theta \Gamma^m
\theta,
\label{Xtrans} \qquad
\delta \bar\theta = \bar \kappa (1+\Gamma),
\end{equation}
and \begin{equation}
\label{rotGamma}
\Gamma = e^{a\over 2} \Gamma'_{(0)} e^{-{a\over 2}}\, ,
\end{equation}
 where
\begin{equation}
a=\cases{+{1\over2} Y_{jk} \gamma^{jk} \Gamma_{11}\qquad\qquad &{\rm
IIA\, ,}\cr
-{1\over2} Y_{jk}   \gamma^{jk} \sigma_3\otimes 1\qquad\qquad &{\rm IIB\, .}}
\end{equation}

Here $\Gamma_{(0)}'$ is the product structure,   independent on the BI field,
$(\Gamma_{(0)}')^2=1 , tr \,
\Gamma_{(0)}'=0$).
 All dependence on BI field ${\cal F} = ``\tan "Y$ is in the exponent
\cite{BKOP}.
The matrix
 $\Gamma_{11}$ in IIA and  $\sigma_3\otimes \ 1$ in IIB theory
anticommute with $\Gamma_{(0)}'$ and with $\Gamma$. Therefore in the basis
where  $\Gamma_{11}$
and $\sigma_3\otimes \ 1$ are
diagonal, $\Gamma_{(0)}'$ and $\Gamma$ are  off-diagonal.
\begin{equation}
\Gamma_{(0)}'  = \pmatrix{0&
\hat  \gamma \cr
\cr               \hat \gamma^{-1} &0}\, .
\end{equation}
We  introduced
a $16\times 16$-dimensional matrix $\hat  \gamma$ which does not depend on BI
field.
Now   the $\kappa$-symmetry generator can be presented in a useful off-diagonal
form
\begin{equation}
\Gamma = \pmatrix{0&
\hat \gamma e^{{\hat a}}\cr
\cr                ( \hat \gamma e^{{\hat a}})^{-1}&0}\, .
\end{equation}
where
\begin{equation}
\hat a=\cases{+{1\over2} Y_{jk} \gamma^{jk} \qquad\qquad &{\rm
IIA\, ,}\cr
-{1\over2} Y_{jk} \gamma^{jk}\qquad\qquad &{\rm IIB\, .}}
\end{equation}

The fact that $\Gamma$ is off-diagonal and that the matrix $\gamma e^{{\hat
a}}$
is invertible is quite important and the significance of this was already
discussed in \cite{Ag1,BKOP}.
In particular this allows us to consider only irreducible $\kappa$-symmetry
transformations by imposing a Lorentz covariant
condition on $ \bar \kappa$ of the form
\begin{eqnarray}
\bar \kappa_1  &=& 0  \qquad  \bar \kappa_2   \neq  0 \qquad {\rm IIA}
\\
\bar \kappa_2 &=& 0  \qquad   \bar \kappa_1  \neq  0 \qquad {\rm IIB}
\end{eqnarray}
In this way we have an irreducible  16-dimensional $\kappa$-symmetry since
the matrix $\hat \gamma e^{\hat a}$ is invertible, acting as
\begin{eqnarray}
\delta \bar \theta_1 &=& \bar \kappa_2 \hat \gamma e^{{\hat a}}  \qquad
\delta \bar \theta_2 = \bar \kappa_2 \hskip 2.2 cm  \delta X^m = - \bar
\kappa_2\Gamma^m \theta_2
 \hskip 2.2 cm  {\rm IIA} \label{kappaA}\\
\delta \bar \theta_1 &=& \bar \kappa_1 \hskip 1.5  cm
\delta \bar \theta_2 = \bar \kappa_1 (\hat \gamma e^{\hat a} )^{-1} \qquad
\delta X^m = - \bar
\kappa_1\Gamma^m \theta_1
\hskip 2.2  cm  {\rm IIB}  \label{kappaB}
\end{eqnarray}

\section{Fermionic constraints prior to gauge-fixing}

The Hamiltonian analysis of supersymmetric extended objects with
$\kappa$-symmetry
requires the knowledge of the fermionic constraints. We split the worldvolume
coordinates $\sigma^\mu$ into time $ \sigma^0= \tau $ and space part $\sigma^a
$
where $a=1, \dots , p.$ In the first approximation we will neglect the
space-dependence on $\sigma^a $ of  spinors $\theta$.

To find fermionic  constraints we observe that our $\kappa$-symmetric D-p-brane
actions depend on the
following combinations of the time derivatives of the fields:
\begin{equation}
L_{\rm DBI} \left ( \dot X^m - \bar \theta_1 \Gamma^m \dot \theta_1 - \bar
\theta_2
\Gamma^m \dot \theta_2, \; \dot A_a - [\bar \theta_1 \Gamma^m \dot \theta_1 -
\bar \theta_2 \Gamma^m \dot \theta_2]\Pi_{ma}\right )
 - L_{\rm WZ} ( \bar  \theta_1  T_{\rm WZ} \dot \theta_2 + c.c.) \ ,
\end{equation}
and $T_{\rm WZ} \equiv \Gamma^A Z_A $ in the WZ term and $\Gamma^A$ are given
by  an
odd number of antisymmetrized
$\Gamma$-matrices in IIA theory and an even number in IIB case.

 We introduce canonical momenta  ${\bf P}_{m}, P^0, P^a, P_{\theta_1},
P_{\theta_2}   $ to $X^m, A_0, A_a , \theta_1, \theta_2$. The fermionic
constraints follow
\begin{eqnarray}
\bar \Phi_1  &=&  \bar P_{\theta_1} +\bar \theta_1 ({\bf P}_{m} + P^a \Pi _a
^m) \Gamma^m + \bar \theta_2 \Gamma^M Z_M \ ,\\
\nonumber\\
\bar \Phi_2  &=&  \bar P_{\theta_2} +\bar \theta_2 ({\bf P}_{m} - P^a \Pi _a
^m) \Gamma^m + \bar \theta_1 \Gamma^M Z_M \ .
\end{eqnarray}
These 32 constraints can be shown to represent 16 first class constraints and
16 second class ones.
The Poisson brackets of these constraints are given by (terms with possible
derivatives of the delta functions $\delta^p( \sigma^a -\tilde  \sigma^a)$ are
omitted)
\begin{eqnarray}
\{ \Phi_1 (\tau , \sigma^a)  ,  \Phi_1 (\tau , \tilde \sigma^a) \}&=& 2 ({\bf
P}_{m} + P^a \Pi _a ^m) C \Gamma^m \delta^p( \sigma^a -\tilde  \sigma^a) +
\dots\\
\nonumber \\
\{ \Phi_2 (\tau , \sigma^a)  ,  \Phi_2 (\tau , \tilde \sigma^a) \}&=& 2 ({\bf
P}_{m} - P^a \Pi _a ^m) C \Gamma^m \delta^p( \sigma^a -\tilde  \sigma^a) +
\dots\\
\nonumber \\
\{ \Phi_1 (\tau , \sigma^a)  ,  \Phi_2 (\tau , \tilde \sigma^a) \}&=& 2
\Gamma^A Z_A \delta^p( \sigma^a -\tilde  \sigma^a) + \dots
\label{mix}\end{eqnarray}
These brackets realize d=10, N=2 supersymmetry algebra  with central
extensions. The R-R charges in this algebra $Z_A$ are due to the structure of
the WZ part of the action. For example, in $p=0$ case we have $\Gamma^A Z_A =
Z$, where $Z$ is the mass of the D-0-particle.
In $p=1$ we get $\Gamma^A Z_A = \Gamma^m  \Pi _\sigma ^m$, for $p=2$ this term
is $\Gamma^A Z_A = \Gamma^{m_1} \Gamma^{m_2}  \epsilon^{ab} [ \Pi _{a m_1}  \Pi
_{ b m_2} + {\cal F}_{ab}] $, $a=1, 2$, etc.
The terms with the $P^a \Pi _a ^m$ could be considered as defining the NS-NS
charge.

The problem of covariant quantization of the fundamental string was the
impossibility to  disentangle these constraints into the first class and the
second class covariantly. For the D-branes the situation is different due to
the presence of the R-R charges  which appear in the bracket of $
\Phi_1$  and $ \Phi_2$ in (\ref{mix}). We will see later that the
reparametrization constraints  in presence of R-R central charges are such that
the bracket for e. g. $ \{ \Phi_1,\,  \Phi_1 \}$ is invertible.

\section {Gauge-fixed $\kappa$-symmetry}

{}From now on we will study the possibilities to quantize these actions. One of
the possibilities which we will pursue here is to immediately gauge-fix
$\kappa$-symmetry by choosing the gauge for theta's. Here again we will follow
\cite{Ag1} and simply take
\begin{eqnarray}
\theta_2 &=& 0 \qquad {\rm IIA}  \qquad \theta_1 \equiv \lambda
\label{thetasA}\\
\theta_1 &=& 0  \qquad {\rm IIB}  \qquad \theta_2 \equiv \lambda
\label{thetasB}
\end{eqnarray}
Note that our choice of irreducible $\kappa$-symmetry (which is not unique) was
made here with the purpose to explicitly eliminate $\theta_2 $ ($\theta_1$) in
IIA (IIB) case using  $\delta \bar \theta_2 = \bar \kappa_2
$ ( $\delta \bar \theta_1 = \bar \kappa_1$).
The gauge-fixed action has one particularly useful property: the Wess-Zumino
term vanishes in this gauge \cite{Ag1}. We are left with the reparametrization
invariant action:
\begin{equation}
S_{\kappa-{\rm fixed}}= - \int d^{p + 1} \sigma \sqrt{- {\rm {\rm det}}
(G_{\mu\nu} + {\cal F}_{\mu\nu})} \ ,
\label{gaugefixed}\end{equation}
\begin{equation}
G_{\mu\nu} = \eta_{mn} \Pi_\mu^m \Pi_\nu^n \ , \qquad \Pi_\mu^m = \partial_\mu
X^m - \bar\lambda \Gamma^m \partial_\mu \lambda \ ,
\end{equation}
\begin{equation}
{\cal F}_{\mu\nu} = [\partial_\mu A_\nu - \bar \lambda \Gamma_m
\partial_{\mu}\lambda
\left(\partial_{\nu} X^m - {1\over 2} \bar
\lambda \Gamma^m \partial_{\nu}\lambda \right)] - (\mu \leftrightarrow \nu) \ .
\end{equation}
The justification of this procedure can be done either by showing that the
remaining action does not have fermionic gauge symmetries anymore (which was
done in \cite{BKOP}) or by the study of the Hamiltonian of the gauge-fixed
theory and the constraints.
In \cite{Ag1} the remaining reparametrization symmetry of the theory was
gauge-fixed by choosing a static gauge. The resulting gauge-fixed action
does not have fermionic degeneracy and this also shows that the choice of a
gauge-fixing made in \cite{Ag1} is acceptable.

We will take from now on a different approach and study the Hamiltonian of the
theory after $\kappa$-symmetry is gauge-fixed.

\section{Hamiltonian structure of the theory}

The set of  canonical momenta and coordinates of the theory in
(\ref{gaugefixed})  includes
$({\bf P}_{m}, X^m) \ , (P^0, A_0) \ , \\
(P^a, A_a) \ ,  (P_{\lambda} , \lambda)$.  All
expressions are relatively simple if as before we neglect in the first
approximation terms with worldvolume space derivatives on spinors ${\partial
\lambda \over \partial
\sigma^a}$. The phase space action (\ref{gaugefixed}) can be brought to the
canonical form
\begin{equation}
L= p_i \partial_{0} q^i + \xi^\alpha t_\alpha(p,q) \ ,
\end{equation}
where $t_\alpha (p,q)$ are some constraints of the first and second kind.
\begin{eqnarray}
L_{\rm can} &&=  {\bf P}_{m}\partial_{0}X^{m}
  +  P^{a}F_{0a} + \bar P_{\lambda} \partial_{0} \lambda \nonumber \\
  &&- \xi ({\bf P}_{m} {\bf P}^{m} + P^{a}G_{ab}P^{b} + T^2\det [ (G+{\cal F}
)_{ab}] ) -\xi_{\rm BI}
P^{0}\nonumber \\
 &&- \xi^{a}\left({\bf P}_{m}  \Pi^{m}_a + P^{b}{\cal F}_{ab}  \right)
+ \left(   \bar P_{\lambda} +\bar \lambda ({\bf P}_{m} + P^a \Pi _a ^m)
\Gamma^m \right) \psi  + \dots .
\end{eqnarray}
Recently a Hamiltonian analysis of the bosonic part of the D-brane action was
performed in \cite{LU}.  Our $\kappa$-symmetry-gauge-fixed action is a
supersymmetric generalization of the theory, studied in \cite{LU}. Indeed we
have
one manifest 16-dimensional global supersymmetry
\begin{equation}
\delta_\epsilon \lambda = \epsilon, \qquad \delta_\epsilon X^m = \bar\epsilon
\Gamma^m
\lambda \ . \label{susytrans1}
\end{equation}
The second 16-dimensional global supersymmetry   exists since the gauge
$\theta_2 =0$ required $\kappa_2 + \epsilon _2=0$ and $\delta \bar \theta_1 = -
\bar \epsilon_2 \hat \gamma e^{\hat a}$  in IIA theory and analogous in IIB
case.
At this stage it is important that we have a manifestly realized 16-dimensional
supersymmetry. This allows us to use the  structure of the bosonic Hamiltonian
theory \cite{LU} and perform an N=1 supersymmetrization of it.
The dots contain terms with space derivatives on spinors ${\partial \lambda
\over \partial
\sigma^a}$. There is also a secondary ``Gauss law" constraint \cite{LU}.
The bosonic constraints related to the reparametrization gauge symmetry, the
time
reparametrization and space reparametrization constraints (with Lagrange
multipliers $\xi, \xi_a$) are
\begin{eqnarray}
t&=&{\bf P}_{m} {\bf P}^{m} + P^{a}G_{ab}P^{b} +
     T^2\det [ (G+{\cal F} )_{ab}] \ , \\
  t_{a}&=&{\bf P}_{m}  \Pi^{m}_a + P^{b}{\cal F}_{ab}\ .
\end{eqnarray}
The constraint related to  the abelian
gauge symmetry  (with the Lagrange multiplier $\, \xi_{\rm BI} $) is
\begin{equation}
t_{\rm BI} = P^{0} \ .
\end{equation}
These constraints are not much different from the constraints in the bosonic
theory \cite{LU}. The new feature here is
the presence of 16 fermionic constraints:
\begin{equation}
\bar \Phi_{\lambda} =  \bar P_{\lambda} +\bar \lambda ({\bf P}_{m} + P^a \Pi _a
^m) \Gamma^m \ .
\end{equation}

If the gauge-fixing
of the previous section is correct, we should find out that the fermionic
constraints are second class, and there is no fermionic gauge symmetry left.
For
this to happen, the Poisson bracket of fermionic constraints has to be
invertible when other constraints are imposed. Thus we have to calculate
\begin{equation}
\{ \Phi_{\lambda }  (\tau , \sigma^a)  ,  \Phi_{\lambda }  (\tau , \tilde
\sigma^a)  \}= ({\bf P}_{m} + P^a \Pi _a ^m) C \Gamma^m \delta^p( \sigma^a
-\tilde  \sigma^a) + \dots
\end{equation}
Here $\dots $ stands for terms with derivatives of the $\delta^p( \sigma^a
-\tilde  \sigma^a)$-function and terms with space derivatives on spinors.
The square of the matrix in the right-hand side of the constraint is
\begin{equation}
[({\bf P}_{m} + P^a \Pi _a ^m) \Gamma^m]^2 = {\bf P}_{m} {\bf P}^{m} +
P^{a}G_{ab}P^{b} + 2 {\bf P}_{m} P^a \Pi _a ^m \ .
\end{equation}
When the reparametrization constraints are imposed, we get
\begin{equation}
([({\bf P}_{m} + P^a \Pi _a ^m) C \Gamma^m]^2)_{T=T_a =0}  = {\bf P}_{m} {\bf
P}^{m} + P^{a}G_{ab}P^{b} = -  T^2\det[ (G+{\cal F} )_{ab}] \ .
\end{equation}
This is quite remarkable:  {\it the invertibility of the second class
constraints}
for D-branes quantized in the {\it Lorentz covariant gauge} relies on 2 basic
facts:
The tension has to be non-vanishing and the space part of the determinant of
the BI action has to be non-vanishing,
\begin{equation}
T\neq 0 \label{tension} \ ,
\end{equation}
\begin{equation}
 \det [ (G+{\cal F} )_{ab}] \neq 0 \ .
\label{det}\end{equation}
To understand the meaning of this restriction on the theory, we may  consider
the  physical states satisfying all first class constraints
\begin{equation}
t|\Psi  \rangle  = t_a | \Psi \rangle  = t_{\rm BI}| \Psi \rangle =0 \ .
\end{equation}
The mass of any such  state we define by the eigenvalue of the square of the
momentum
\begin{equation}
M^2 |\Psi  \rangle  = - {\bf P}_{m} {\bf P}^{m}  | \Psi  \rangle  = \left(
P^{a}G_{ab}P^{b} +
     T^2\det [ (G+{\cal F} )_{ab}] \right)  |\Psi   \rangle \ .
\label{mass}\end{equation}
The mass of any physical state of the theory therefore consists of two positive
contributions: the first and the second term in eq. (\ref{mass}). For
the fermionic constraints to be second class this requirement means that all
physical states of the theory have to have strictly non-vanishing  mass.
\begin{equation}
 M^2 |\Psi   \rangle  \;   \geq
 T^2\det [ (G+{\cal F} )_{ab}]  | \Psi   \rangle     \neq 0 |
\Psi   \rangle \ .
\label{massineq}\end{equation}
 This applies equally well to the ground state of the theory. Thus the
Hamiltonian analysis of the reason why D-branes admit Lorentz covariant
gauge-fixing of $\kappa$-symmetry leads to the following conclusion: As long as
there are mo massless states in the theory, $\kappa$-symmetry admits covariant
gauges.

The existence of R-R charges or equivalently the existence of central
extensions in supersymmetry algebra (\ref{mix}) is related to $M^2 \neq0$
condition ($
T^2\det[ (G+{\cal F} )_{ab}] \neq 0$) as follows. When $T\neq 0$ and when
$\Pi_{am}\neq 0$ and/or
${\cal F}\neq 0$, we have simultaneously the R-R charges in the supersymmetry
algebra and
strictly non-vanishing mass of all physical states in the theory. If $T=0$ or
if
$\Pi_{am}= 0$ and
${\cal F}= 0$, there are no R-R charges and the massless state is not excluded.
However, in this case the covariant gauge is not acceptable, since the
fermionic
constraints are not invertible.

We may compare this Hamiltonian result with the gauge-fixing of D-branes in
static gauges $X^\mu = \sigma^\mu $ which was used  in   \cite{Ag1}. The
tension was
equal one there,  $T=1$. In static gauges
\begin{equation}
G_{\mu\nu} = \eta_{\mu\nu} + \dots \ , \qquad \det (G+ {\cal F})_{ab}  = (\det
\eta_{ab} + \dots)  \neq 0 \ ,
\end{equation}
and therefore both conditions for invertibility of the bracket of covariant
fermionic constraints (\ref{tension}) and (\ref{det}) are satisfied.

\section{Covariant quantization of D-0-brane}

Consider the $\kappa$-symmetric action of a D-0-brane. D-0-brane action does
not
have Born-Infeld field since there is no place for an antisymmetric tensor of
rank 2 in one-dimensional theory.
 The action (\ref{action}) for $p=0$ case reduces to
\begin{equation}
S = -T \left (  \int d\tau \sqrt { - G_{\tau\tau}}
  +  \int  \bar \theta \Gamma^{11} \dot \theta \right) \ .
\label{0action}\end{equation}
This action can be derived from the action of the massless 11-dimensional
superparticle.
\begin{equation}
S = \int d\tau \sqrt {g_{\tau\tau}}  g^{\tau\tau} \left( \dot X^{\hat m}  -
\bar \theta \Gamma^{\hat m} \dot \theta \right)^2 \ , \qquad \hat m =
 0,1,\cdots , 8,9,10.
\label{11}\end{equation}
We may solve  equation of motion for $X^{ \hat {10}}$    as ${\bf P}_{ \hat
{10}} =Z$, where $Z$ is a constant, and use   $ \Gamma^{11} =
\Gamma^{\hat {10}}
$. From this one can deduce a
 first order action
\begin{equation}
S = \int d\tau \left( {\bf P}_{m} (\dot X^m - \bar \theta \Gamma^m \dot \theta)
+ {1\over 2} V (
{\bf P}^2 + Z^2) - Z \bar \theta \Gamma^{11} \dot \theta + \bar \chi_1
d_2\right) \ .
\label{first}\end{equation}
We will show now that the D-0-brane action can be obtained from this one upon
solving equations of motion for ${\bf P}_{m}$, $V, $  $\chi_1$, and $d_2$.
Here  $V(\tau) $ is a Lagrange multiplier, $Z=T$  is some constant parameter in
front of the WZ term and  ${\bf P}^2 \equiv {\bf P}^{m}\eta_{mn} {\bf P}^{n}$.
 The chiral spinors $\chi_1$ and $d_2$ are auxiliary fields. They are
introduced to close the gauge symmetry algebra off shell. To verify that this
first
order action is one of the D-p-brane family actions given in (\ref{action}) we
can use equations of motion for ${\bf P}_{m}$
\begin{equation}
{\bf P}_{m}= -{1\over V} (\dot X^m - \bar \theta \Gamma^m \dot \theta) \ ,
\end{equation}
and for the auxiliary fields $ \chi_1=0$ and $  d_2=0$. The action
(\ref{first})
becomes
\begin{equation}
S = \int d\tau \left( -{1\over 2 V} (\dot X^m - \bar \theta \Gamma^m \dot
\theta)^2 +{1\over 2} V
 Z^2 - Z \bar \theta \Gamma^{11} \dot \theta \right) \ .
\label{inter}\end{equation}
Equation of motion for $V$ is
\begin{equation}
 V^2=- {1\over Z^2 } (\dot X^m - \bar \theta \Gamma^m \dot \theta)^2 \ ,
\end{equation}
and we can insert $ V= -{1\over Z }\sqrt { - (\dot X^m - \bar \theta \Gamma^m
\dot \theta)^2}
$ back into the action (\ref{inter}) and get
\begin{equation}
S = - Z \left(  \int d\tau \sqrt { - (\dot X^m - \bar \theta \Gamma^m \dot
\theta)^2}
  + Z \bar \theta \Gamma^{11} \dot \theta \right) \ .
\end{equation}
This is the action (\ref{action}) for D-0-brane at $T=Z$ as given in
(\ref{0action}).

The action  (\ref{first}) is invariant under the 16-dimensional irreducible
$\kappa$-symmetry and under the
reparametrization symmetry. The gauge symmetries are (we denote $\Gamma^m {\bf
P}_{m}= \sl$):
\begin{eqnarray}
\delta \bar \theta&=& \bar  \kappa_2 ( \Gamma^{11} Z + \sl) \ , \\
\delta X^m &=& -\eta {\bf P}^{m}  - \delta \bar \theta  \Gamma^m \theta - \bar
\kappa_2 \Gamma^m d \ , \\
\delta V &=& \dot \eta + 4 \bar \kappa_2 \dot \theta + 2 \bar \chi_1 \kappa_2\
, \\
\delta \bar \chi &=& \bar \kappa_2   \dot  \sl \ ,\\
\delta d &=&[ {\bf P}^2 + Z^2] \kappa_2 \ .
\end{eqnarray}
Here $\eta(\tau) $ is the time reparametrization gauge parameter and
$\kappa_2(\tau)
=
{1\over 2} (1- \Gamma^{11} ) \kappa (\tau) $ is the 16-dimensional parameter of
$\kappa$-symmetry. The gauge symmetries form a closed algebra
\begin{equation}
[\delta({\kappa_2}) ,\delta({\kappa'_2}) ] = \delta ({\eta = 2 \bar \kappa_2
\sl \kappa'_2} ) \ .
\end{equation}

To bring the theory to the canonical form we introduce  canonical momenta to
$\theta$ and to $V$ and find, excluding auxiliary fields
\begin{equation}
L =  {\bf P}_{m} \dot X^m+ P_V \dot V + \bar P_{\theta } \dot \theta  + {1\over
2} V ({\bf P}^2
 + Z^2)
  +   P_V \varphi + \left (\bar P_{\theta }  + \bar \theta  (\sl + Z
\Gamma^{11})\right ) \psi \ .
\end{equation}

We have  primary constraints $\bar \Phi  \equiv  \bar P_{\theta }  + \bar
\theta  (\sl + Z \Gamma^{11}) \approx 0$
and $P_V=0$. The Poisson brackets for 32 fermionic constraints are
\begin{equation}
\{ \Phi  ,  \Phi  \}=  2 C (\sl  + \Gamma_{11} Z ) \ .
\end{equation}
We also have to require that the constraints are consistent with the time
evolution $\{P_V, H\}=0$. This generates a secondary constraint
\begin{equation}
t=  {\bf P}^2 + Z^2 \ .
\end{equation}
Thus the Hamiltonian is weakly zero and any physical state of the system
satisfying the reparamet-\\
rization constraint is a BPS state $M=|Z|$ since
\begin{equation}
 {\bf P}^2 + Z^2 | \Psi \rangle  = 0 \qquad  \Longrightarrow  \qquad Z^2   |
\Psi \rangle =-  {\bf P}^2  | \Psi \rangle = M^2  |  \Psi  \rangle \ .
\end{equation}
 The $32\times 32$ -dimensional matrix $C (\sl + \Gamma_{11} Z) $ is not
invertible since it squares to zero when the reparametrization constraint is
imposed. This is a reminder of the fact that D-0-brane is a d=11 massless
superparticle.
The 32 dimensional fermionic constraint has a 16-dimensional part which forms a
first class constraint and another 16-dimensional part which forms a second
class constraint.
We notice that the Poisson brackets reproduce the $d=10$, $N=2$ algebra with
the
central charge which can also be understood as $d=11$, $N=1$ supersymmetry
algebra with the
constant value of ${\bf P}_{11}=Z$.

We proceed with the quantization and gauge-fix
$\kappa$-symmetry covariantly   by taking $\theta_2=0, \theta_1 \equiv \lambda
$
and find
\begin{equation}
L^\kappa_{g.f.}= {\bf P}_{m} (\dot X^m - \bar \lambda  \Gamma^m \dot \lambda )
+ {1\over 2} V (
{\bf P}^2 + Z^2) \ .
\end{equation}
The 16-dimensional fermionic constraint
\begin{equation}
 \bar \Phi_{\lambda }  \equiv  (\bar P_{\lambda }  + \bar \lambda  \sl )\approx
0
\end{equation}
forms the Poisson bracket
\begin{equation}
\{  \Phi^{\alpha }_{\lambda }  ,  \Phi ^{\beta }_{\lambda } \}=   2  (\sl
C)^{\alpha \beta}   \label{Pois} \ .
 \end{equation}
The matrix $ \sl C
$ is perfectly invertible as long as the central charge $Z$ is not vanishing.
The inverse to (\ref{Pois}) is
\begin{equation}
\{  \Phi^{\alpha }  ,  \Phi ^{\beta } \}^{-1}  \mid_{t=0} \;
= [2  ( \sl C)^{\alpha \beta}   ]^{-1} = {(C \sl )_{\alpha \beta}
 \over 2  {\bf P}^2} \ .
  \end{equation}

This proves that the fermionic constraints are second class and that the
fermionic part of the Lagrangian
\begin{equation}
- \bar \lambda  \sl \dot \lambda \equiv - i \lambda^\alpha \Phi_{\alpha \beta}
\dot \lambda^\beta \ ,  \qquad \Phi_{\alpha \beta} =-i (C\sl )_{\alpha \beta} \
,
\end{equation}
is not degenerate in a Lorentz covariant gauge. None of this would be true for
a vanishing central charge. Note that in the rest frame ${\bf P}_{0}=M , \vec
{\bf P}=0
$, hence
\begin{equation}
 \Phi_{\alpha \beta} = M \delta_{\alpha \beta} \ .
\end{equation}
For D-0-brane one can covariantly gauge-fix the reparametrization symmetry by
choosing the $V=1$ gauge and including the anticommuting reparametrization
ghosts
$b,c$. This brings us to the following form of the action:
\begin{equation}
L^{\kappa, \eta} _{g.f.}= {\bf P}_{m} \dot X^m - \bar \lambda  \sl \dot \lambda
 + {1\over 2}  (
{\bf P}^2 + Z^2)
 +  b\dot c \ .
\end{equation}
Now we can define Dirac brackets
\begin{eqnarray}
&& \{  \lambda  ,  \bar \lambda \}^*= \{  \lambda , \bar \Phi \}  \{ \bar \Phi_
,  \Phi  \}^{-1}  \{  \Phi  \ , \bar \lambda \} = {   \sl \over 2 {\bf P}^2} =-
{ \sl \over 2 Z^2} \ .
\end{eqnarray}
The generator of the 32-dimensional supersymmetry is
\begin{equation}
\bar \epsilon Q = \bar \epsilon ( \sl  + \Gamma^{11} Z) \lambda \ .
\end{equation}
It forms the following Dirac bracket
\begin{equation}
[\bar \epsilon Q \ , \bar Q \epsilon' ]^* = \bar \epsilon (\sl +
\Gamma^{11} Z)  {   \sl \over 2 {\bf P}^2} (\sl +
\Gamma^{11} Z)  \epsilon'=
\bar \epsilon \Gamma^{\hat m}  {\bf P}_{\hat m}   \epsilon'   = \bar \epsilon
(\sl +
\Gamma^{11} Z)  \epsilon' \ .
\label{diracsusy}\end{equation}
We can also rewrite it in d=11 Lorentz covariant form
\begin{equation}
[\bar \epsilon Q \ , \bar Q \epsilon' ]^*=\bar \epsilon \Gamma^{\hat m}  {\bf
P}_{\hat m}   \epsilon' = \bar \epsilon \; \hat {\sl } \epsilon'\ ,
 \qquad \hat m =
 0,1,\cdots , 8,9,10, \qquad   Z= {\bf P}_{ \hat {10}}\ ,  \qquad \Gamma^{11} =
\Gamma^{\hat {10}} \ .
\label{diracsusy11}\end{equation}
This Dirac bracket realizes the d=11, N=1 supersymmetry algebra
 or, equivalently, d=10, N=2 supersymmetry algebra with the central charge
$Z$.

One can also  to take into account that the path integral in presence of second
class constraints has an additional term with $\sqrt{{\rm Ber} \{\Phi _\lambda
,
\Phi_\lambda  \}} \sim  \sqrt{ {\rm Ber}\, \Phi_{\alpha \beta} }$~
\cite{Fradkin}, see Appendix. It can be
used to make a change of variables
\begin{equation}
S_\alpha  = \Phi^{1/2}_{\alpha \beta}    \; \lambda^\beta \ .
\end{equation}
The action becomes
\begin{equation}
L= {\bf P}_{m} \dot X^m-  i S_\alpha  \dot S_\alpha   +  b\dot c - H
\end{equation}
\begin{equation}
H= - {1\over 2} (
{\bf P}^2 + Z^2) \ .
\end{equation}
The generators of global supersymmetry commuting with the Hamiltonian take the
form
\begin{equation}
\bar  \epsilon  Q    =   \bar \epsilon ( \sl  + \Gamma^{11} Z)  \Phi^{-1/2} S \
{}.
\end{equation}
Taking into account that $ \{ S _\alpha  ,  S_\beta  \}^* =-{i\over 2}
\delta_{\alpha \beta}$
we have again realized $d=10$, $N=2$  supersymmetry algebra in the form
(\ref{diracsusy}) or  (\ref{diracsusy11}).

The terms with anticommuting fields $S_\alpha $ can be rewritten in a form
where it is clear that they can be interpreted as   world-line spinors,
 \begin{equation}
L= {\bf P}_{m}  \partial_0 X^m +  \bar S_\alpha  \rho^0 \partial_0  S_\alpha
+  b\dot c - H \ .
\end{equation}
Here $\bar S_\alpha =  i S_\alpha  \rho^0$ and $(\rho^0)^2 =-1$, $\rho^0=i$
being a 1-dimensional matrix.

Thus, we have the original 10 coordinates $X^m$ and their conjugate momenta
${\bf P}_{m}$, and a pair of reparametrization ghosts. There are also 16
anticommuting world-line spinors $S$, describing 8 fermionic degrees of
freedom. The Hamiltonian  is quadratic.
The  ground state with   $M^2=Z^2$ is the state with the minimal value of the
Hamiltonian.
Thus for the D-superparticle one can see that the condition for the covariant
quantization is satisfied in the presence of a central charge which makes the
mass of a physical state non-vanishing. The global supersymmetry algebra is
realized in a covariant way, as different from the light-cone gauge.

 \section {Conclusion}

Thus, we have confirmed here the conclusion of the work \cite{Ag1} that one can
covariantly quantize  D-p-branes. However, as different from  \cite{Ag1}, we
did not use  the static gauge for fixing the reparametrization symmetry, and
analysed the Hamiltonian structure of the theory. This   allowed us to clarify
the reason and the generic condition under which a covariant quantization of
D-p-branes is possible: the ground state has to be strictly massive, $M^2_{\rm
groundstate}=Z^2 >0$. Technically, this reduces to  2 conditions:

i) The tension of the D-p-brane $T$ has to be non-vanishing.

ii)  $\det (G+{ \cal F})_{ab}$  has to be non-vanishing  ($a,b$ are space
components of the brane).

Those two conditions for the D-p-brane are basically equivalent to the
definition of this object. Both these conditions have to be satisfied if there
are non-vanishing central charges in the supersymmetry algebra which are due to
the existence of R-R charges of the D-p-brane. The cross term in the left-right
part of N=2 supersymmetry algebra
\begin{equation}
\{ Q_\alpha ,  \; \tilde Q_\beta \} =2   \sum_A  (C \Gamma^{A})_{\alpha
\beta} T {Q_{A}^R \over p!}
\end{equation}
does not vanish when conditions\, i) and  ii)  for covariant quantization are
satisfied. These conditions provide the {\it invertibility of  the second class
fermionic constraints in Lorentz covariant gauges}.

 As the special case we have performed a covariant quantization of D-0-brane.
The resulting supersymmetry generator is $d=10$ Lorentz covariant and the Dirac
bracket of the quantized theory form $d=10$, $N=2$ supersymmetry algebra with
a central charge.

By announcing that  D1-brane can be covariantly quantized we have to be able to
deal covariantly with the  IIB Green-Schwarz action since it is SL(2,R) dual to
the D1-brane.  Our conclusion is that this is indeed possible under the
condition that the massless state is projected out from the fundamental IIB
string. In \cite{Ag1} this was effectively demonstrated since the static gauge
used there corresponds to the IIB superstring wrapped around the circle and
such object does not  have massless excitations. Thus, we conclude that the old
problem of impossibility to covariantly quantize the fundamental string can be
avoided for IIB fundamental string for all states but massless. The dual
partner of it,  D-1 string, does not have massless states, and duality  works
only in the sector of massive excitations of these two theories, quantized
covariantly. To confirm this picture and better understand these issues  one
has to quantize these two dual theories in the conformal gauge for the
reparametrization symmetry and in covariant gauge for
$\kappa$-symmetry.

One has to note here that the choice of  the Lorentz covariant fermionic gauge
is not trivial for objects dual to D-branes. It has been explained in
\cite{BKOP} that the projector $\Gamma$  of $\kappa$-symmetry of D-p-branes
anticommutes with  Lorentz covariant chiral projectors in d=10. This
technically explains why
Lorentz covariant gauges are capable of removing the degeneracy of the theory
due to  $\kappa$-symmetry on D-p-branes.
This is not the case for the fundamental strings, and the choice of the
acceptable Lorentz covariant fermionic gauge has to be done properly. It is
important that in IIB theory with two chiral spinors any gauge of the type
\begin{equation}
\theta_1 = c \theta_2 \ ,
\end{equation}
where $c$ is and arbitrary constant, is consistent with Lorentz symmetry.

Similar observations   apply to the D2-brane
 versus the eleven-dimensional supermembrane \cite{BST}  which are also
 related to each other, this time via duality on the worldvolume.
Is it possible that we can covariantly quantize the D2-brane but not the
dual d=11 supermembrane? First notice that the chiral projectors ${1\over 2}
(1\pm \Gamma^{11})$ which have been used for the covariant quantization of a
D-2-brane  are covariant in d=10 but not Lorentz covariant in d=11. Secondly,
our requirement of the absence of massless states in d=10 for the D-2-brane
does not exclude the massless d=11 state since we have a BPS ground state in
d=10. As we have seen in quantization of the D-0-brane in Sec. 7, the BPS
condition $- {\bf P}_{10}^2= M^2_{10}=Z^2$ is in fact equivalent to the
statement that the 11-dimensional superparticle is massless, $- {\bf P}_{10}^2-
Z^2= - {\bf P}_{11}^2=0$.

At this stage we have learned that the 10d Lorentz covariant quantization of
the D-2-brane may be a step towards understanding of the spectrum of states of
the fundamental theory. In particular, we have found the mass formula for the
D-2-brane
(\ref{massD2}) which is a d=10 Lorentz covariant generalization of the mass
formula of the supermembrane quantized in the light-cone gauge
\cite{BSTa,DWHN}.

\section*{Acknowledgements}
 I had stimulating discussions with Eric Bergshoeff, David Gross, Gary
Horowitz, Joe Polchinsky, Joachim Rahmfeld, and Wing Kai Wong. This work   is
supported by the NSF grant PHY-9219345.

\section*{Appendix: Quantization in canonical gauges}

Quantization of an arbitrary Bose-Fermi-system with the first and second class
constraints in canonical gauges was performed by E. Fradkin and collaborators
\cite{Fradkin}. One starts with the classical Lagrangian of the form
\begin{equation}
L= p_i \dot q^i - H_0(p,q) - \xi^k \Theta_k (p,q) - \xi^\mu T_\mu (p,q) \ .
\end{equation}
The  first class constraints obey the relations
\begin{equation}
\{T_\mu , T_\nu \} \mid_{T=0, \Theta=0} =0\ , \qquad \{H_0 \ , T_\mu
\}\mid_{T=0, \Theta=0}=0\ ,
\end{equation}
and for the second class constraints we have
\begin{equation}
{\rm Ber}\{\Theta_k , \Theta_l \} \mid_{T=0, \Theta=0} \neq 0 \ .
\end{equation}
The symbol $\{ \}$ stands for the Fermi-Bose extension of the Poisson bracket
and ${\rm Ber}$ (Berezinian) is a superdeterminant of the matrix  $\{\Theta_k ,
\Theta_l \} \equiv \Theta_{kl}$.
The Dirac bracket is defined as
\begin{equation} \ .
\{A , B \}^* = \{A , B \} - \{A , \Theta_k \} (\Theta_{kl})^{-1}  \{\Theta_l ,
B \} \ .
\end{equation}
The path integral in canonical gauges $\Phi^\mu (p,q)=0$ (where all ghosts are
non-propagating fields)
 is given by
\begin{equation}
Z = \int \exp \{ i S[q,p, \xi, \pi]\} \prod{({\rm Ber}\{\Phi  , T \}^* \delta
(\Theta) ({\rm Ber}\{\Theta , \Theta\})^{1/2} dq \; dp\; d\xi d\pi}) \ ,
\end{equation}
where the action is
\begin{equation}
S=\int (p_i \dot q^i - H_0(p,q)  - \xi^\mu T_\mu (p,q) - \pi _\mu \Phi^\mu)
d\tau \ .
\end{equation}

\end{document}